\documentclass{optica-article}
\journal{opticajournal}
\articletype{Research Article}
\usepackage{lineno}
\usepackage{nicematrix}
\usepackage{adjustbox}
\usepackage{arydshln}
\usepackage{float}
\begin{document}
\title{Probabilistic volumetric speckle suppression in OCT using deep learning}
\author{Bhaskara Rao Chintada,\authormark{1,2,*} Sebasti\'an Ruiz-Lopera,\authormark{1,3} Ren\'e Restrepo,\authormark{4} Brett E. Bouma,\authormark{1,2,5} Martin Villiger,\authormark{1,2} and N\'estor Uribe-Patarroyo\authormark{1,2}}
\address{\authormark{1}Wellman Center for Photomedicine, Massachusetts General Hospital, Boston, MA 02114, USA\\
\authormark{2}Harvard Medical School, Boston, MA 02115, USA\\
\authormark{3}Department of Electrical Engineering and Computer Science, Massachusetts Institute of Technology, Cambridge, MA 02142, USA\\
\authormark{4}Applied Optics Group, Universidad EAFIT, Carrera 49 \# 7 Sur-50, Medellín, Colombia\\
\authormark{5}Institute of Medical Engineering and Science, Massachusetts Institute of Technology, Cambridge, MA 02142, USA}
\email{\authormark{*}bchintada@mgh.harvard.edu}%

\begin{abstract}
We present a deep learning framework for volumetric speckle reduction in optical coherence tomography (OCT) based on a conditional generative adversarial network (cGAN) that leverages the volumetric nature of OCT data.
In order to utilize the volumetric nature of OCT data, our network takes partial OCT volumes as input, resulting in artifact-free despeckled volumes that exhibit excellent speckle reduction and resolution preservation in all three dimensions.
Furthermore, we address the ongoing challenge of generating ground truth data for supervised speckle suppression deep learning frameworks by using volumetric non-local means despeckling-TNode to generate training data.
We show that, while TNode processing is computationally demanding, it serves as a convenient, accessible gold-standard source for training data; our cGAN replicates efficient suppression of speckle while preserving tissue structures with dimensions approaching the system resolution of non-local means despeckling while being two orders of magnitude faster than TNode.
We demonstrate fast, effective, and high-quality despeckling of the proposed network in different tissue types acquired with three different OCT systems compared to existing deep learning methods.
The open-source nature of our work facilitates re-training and deployment in any OCT system with an all-software implementation, working around the challenge of generating high-quality, speckle-free training data.
\end{abstract}

\section{Introduction}
Optical coherence tomography (OCT) is a cross-sectional optical imaging technique that provides high-resolution images of biological tissue~\cite{huang1991optical} and has become a well-established clinical diagnostic imaging tool in ophthalmology. 
Due to the coherent nature of OCT, tomograms contain speckle, which degrades the image quality and hinders visual interpretation~\cite{schmitt1999speckle, bashkansky2000statistics, karamata2005speckle, goodman2007speckle}.
Speckle reduction has been an active topic of interest in OCT community and a plethora of techniques have been developed in the literature, which can be broadly classified as hardware-based~\cite{iftimia2003speckle, pircher2003speckle, desjardins2007angle, alonso2011speckle, kennedy2010speckle} and signal-processing methods~\cite{ozcan2007speckle, gargesha2008denoising, jian2009speckle, wong2010general, jian2010three, fang2012sparsity, szkulmowski2012efficient, wang2012adaptive, szkulmowski2013averaging, yin2013speckle, chong2013speckle, aum2015effective, cheng2016speckle, cuartas2018volumetric}. 
Hardware-based methods have the potential to produce higher-quality images.
However, hardware modifications and different data acquisition strategies make them too complex for broad adoption and incompatible with imaging \emph{in vivo}.
For instance, angular compounding requires long acquisition times and the sample must be static during the entire acquisition; modulation of the point spread function (PSF) of the illumination beam carries a signal-to-noise ratio (SNR) penalty as well as increased acquisition times~\cite{iftimia2003speckle, pircher2003speckle, desjardins2007angle, alonso2011speckle, kennedy2010speckle}.
Cuartas-V\'elez, et al, have discussed in more detail the merits and limitations of the hardware-based and signal-processing methods for speckle reduction in our previous work~\cite{cuartas2018volumetric}.
Among signal-processing methods, the non-local probabilistic despeckling method TNode~\cite{cuartas2018volumetric} exploits volumetric information in OCT tomograms to estimate the incoherent intensity value at each voxel.
TNode efficiently suppresses speckle contrast while preserving tissue structures with dimensions approaching the system resolution. It is, however, computationally very expensive, a common problem with exhaustive search methods: processing a typical retinal OCT volume takes a few hours.

Deep learning methods have been explored for speckle mitigation by posing the task as an image-to-image translation problem~\cite{ma2018speckle, halupka2018retinal, dong2020optical, guo2020unsupervised, kande2020siamesegan, zhou2021speckle, wang2021semi, shi2019despecnet, abbasi2019three, devalla2019deep, bobrow2019deeplsr, qiu2020noise, menon2020novel, gour2020speckle, gisbert2020self, apostolopoulos2020automatically, varadarajan2022novel, hu2020retinal, oguz2020self, rico2022real, ni2022hybrid}.
The main goal of an image-to-image translation network is to learn the mapping between an input image and an output image~\cite{isola2017image,pang2021image} either through a supervised modality using training pairs of input and target images, or through an unsupervised modality using independent sets of input images and target images when paired examples are scarce or not easy to generate.
Supervised methods are known to yield better results as the model has access to the ground truth information during training.
Previous despeckling efforts using deep learning include the use of variants of generative adversarial networks (GANs)~\cite{ma2018speckle, halupka2018retinal, dong2020optical, guo2020unsupervised, kande2020siamesegan, zhou2021speckle, wang2021semi}, variants of convolutional neural networks (CNN) based methods~\cite{shi2019despecnet, abbasi2019three, devalla2019deep, qiu2020noise, menon2020novel, gour2020speckle, gisbert2020self, apostolopoulos2020automatically,varadarajan2022novel} and fusion networks~\cite{hu2020retinal,  oguz2020self,rico2022real,ni2022hybrid}.
However, with few exceptions~\cite{ni2021sm}, these methods have been based on the compounding of multiple B-scans, acquired at the same sample location, to generate ground truth despeckled tomograms. This presents significant limitations: B-scan compounding does not reduce speckle contrast unless the component images capture a variation of
microstructural organization within the sample, a condition that is typically only satisfied inside blood vessels. We note that inaccuracies in scanning or motion artifacts may also be leveraged to provide data for B-scan compounding, but these approaches induce a direct penalty to spatial resolution.
In addition, \emph{all these methods have focused on two-dimensional speckle suppression}; their performance on volumetric tomograms has not been demonstrated. We expect that any three-dimensional manipulation of processed tomograms (i.e., \textit{en face} projections) will contain artifacts due to the aforementioned B-scan-wise processing, disrupting the continuity of tissue structures along the slow-scan axis.
In addition, we argue that speckle suppression based on two-dimensional data cannot provide the neural network with complete information on volumetric structures in the training data, and thus it is expected to perform poorly on structures that have a small cross section in a given B-scan.

To overcome the limitations in the state of the art, we present a workflow to utilize volumetric information present in OCT data for near-real-time speckle suppression and enable high-quality deep-learning based volumetric speckle suppression in OCT. We exploit ground-truth training data generated using the non-local probabilistic despeckling method, TNode~\cite{cuartas2018volumetric} and our neural network uses a new cGAN that receives OCT partial volumes as inputs, exploiting three-dimensional structural information for speckle mitigation.
Our hybrid deep-learning--TNode-3D (DL-TNode-3D) enables easy training and implementation in a multitude of OCT systems without relying on specialty-acquired training data. 

\section{Methods}
\subsection{Data}
In this study, we trained our network on three different custom-built frequency domain OCT systems; one system used a wavelength-swept light source (Axsun Technologies, Inc., MA, USA) having a spectral bandwidth of 91~nm centered at 1040~nm and a sweep repetition rate of 100~kHz. This system was integrated with the ophthalmic interface of a Spectralis OCT device (Heidelberg Engineering, Germany), which provided eye tracking and fixation capabilities and a transverse resolution (defined hereafter as $e^{-2}$ focal spot diameter after considering the confocal effect) of 18~\textmu m~\cite{braaf2018complex}.
The second system operated with a vertical-cavity surface-emitting laser (VCSEL) that provided a spectral bandwidth of 90~nm centered at 1300~nm wavelength and 100~kHz sweep repetition rate~\cite{cannon2021layer}. This system had a transverse resolution of 12~\textmu m.
The third system was based on a custom-made wavelength-swept laser that utilized a semiconductor optical amplifier (Covega Corp., BOA-4379) as the gain medium and a polygon mirror scanner (Lincoln Laser Co.) as the tunable filter to rapidly sweep the wavelength at a rate of 54~kHz.
This laser had a center wavelength of 1300~nm and bandwidth of 110~nm~\cite{ren2017label} and the system provided a transverse resolution of 16.5~\textmu m.
Herein, we refer to these systems as ophthalmic, VCSEL, and polygon systems, respectively.
All the datasets used in this study were acquired using Nyquist sampling in the fast and slow scan axes.
All systems were provided with polarization-diverse balanced detection; thus, all A-lines were recorded for two detection polarization channels.

We selected a volume of interest (VOI) in each dataset consisting of a varying number of B-scans, A-lines per B-scan, and depth samples per A-line.
We trained our network for each system separately. 
For the ophthalmic system, the network was trained and tested using different regions of the retina from two healthy subjects, whereas for VCSEL and polygon systems, we trained the network with different tissue types \textit{ex vivo} and \textit{in vivo}, and tested using an additional tissue type not used in training.
The tissue types used for training and testing for ophthalmic, VCSEL and polygon systems are tabulated in Table~\ref{tab:datasets}.
In this study, we have used 5 OCT volumes for training and testing for each system (3 $\times$ 5 = 15 volumes in total).
\begin{table}
    \begin{center}
    \begin{adjustbox}{width=\textwidth}
    \begin{NiceTabular}{|c|c|c|} \hline
        \textbf{System} & \textbf{Training} & \textbf{Testing} \\ \hline 
         Ophthalmic~\cite{braaf2018complex} & Retina ($\times$3) & Retina ($\times$2)\\ 
         VCSEL~\cite{cannon2021layer} & Chicken heart ($\times$1) and leg ($\times$1), nail bed ($\times$1) & Ventral($\times$1) and dorsal ($\times$1) finger skins\\ 
         Polygon~\cite{ren2017label} & Chicken leg ($\times$2), dorsal finger skin ($\times$1) & Nail bed ($\times$2)\\\hline
    \end{NiceTabular}
    \end{adjustbox}
    \end{center}
    \caption{Overview of datasets used for training and testing for different OCT systems in this study.}
    \label{tab:datasets}
\end{table}
\subsection{Tomogram pre-processing}\label{sec:pre-processing}
Using MATLAB (MathWorks, USA), the acquired OCT fringes were mapped to a linear wavenumber space, numerically compensated for dispersion, apodized with a Hanning window, zero-padded to the next power of two, and Fourier transformed to reconstruct raw complex-valued tomograms with a final pixel size in the axial direction of 4.8~\textmu m, 5.3~\textmu m and 5.8~\textmu m for the ophthalmic, VCSEL, and polygon systems, respectively, assuming unity index of refraction. 
Data from the VCSEL system was acquired using the $k$-clock from the light source, therefore, it did not require linearization in wavenumber space.

Data acquired \textit{in vivo} further required inter-B-scan bulk-motion correction to preserve the continuity of tissue along the slow axis direction and enable volumetric despeckling. The reconstructed complex-valued tomograms were phase-stabilized along the fast scan axis and then low-pass filtered with the optimum filter~\cite{ruiz2020computational}. Phase-stabilization and filtering, applied to each polarization detection channel, enabled the use of efficient sub-pixel image registration~\cite{guizar2008efficient}, which we used to determine and correct for axial and lateral sub-pixel shifts between adjacent B-scans. After these steps, we calculated the tomogram intensity as the sum of the intensity (squared absolute value) of each polarization detection channel. We then saved the tomogram intensity in logarithmic scale in single precision format for further processing with TensorFlow library in Python.

\subsection{Non-local-means despeckling (TNode)}
Tomographic non-local-means despeckling (TNode)~\cite{cuartas2018volumetric} exploits volumetric information in OCT by making use of 3D similarity windows to retrieve the weights from the volumetric patch-similarity. 
This method uses a 3D search window, which consists of depth, $z$, and both fast-axis, $x$, and slow-axis, $y$, information.
For tomograms acquired \textit{in vivo}, motion correction was applied to guarantee continuity of tissue structures along the slow axis (see Sec.~\ref{sec:pre-processing}).
TNode efficiently suppresses speckle while preserving tissue structures with dimensions approaching the system resolution.
Despite the merits of this method, it is very computationally demanding; even after making it more computationally efficient by an order of magnitude compared to our original implementation (see Supplement 1 and \href{https://github.com/bhaskarachintada/DLTNode.git}{Code~1},~Ref.~\cite{Chintada2023DLTNode}).

In this study, a search window of size $2 \times 8+1$ was chosen along the slow-scan axis for TNode speckle reduction, and thus the network was defined to accept partial volumes with $n = 8$; the network is easily modified to accept the partial volumes with any desired $n$.
We set the base filtering parameter, $h_0 = 80\times 10^{-3}$ and the SNR-dependent parameter, $h_1 = 40\times 10^{-3}$ to process the OCT volumes. 

\subsection{cGANs for volumetric speckle suppression}
To suppress the speckle in a cross-section $x_i$, several conventional signal-processing methods in the literature use information from its $2n$ neighborhood cross-sections from $x_{i-n}$ to $x_{i+n}$ in an OCT volume $V$.
We denote the cross sections from $x_{i-n}$ to $x_{i+n}$ as partial volume $v_i$, and $y_i$ as the speckle-suppressed cross-section corresponding to $x_i$.
Mapping $x_i$ to ${y_i}$ can be treated as an image-to-image translation problem.
cGANs are known to perform extremely well in these problems compared to other CNN-based methods~\cite{isola2017image}.
cGANs can be adapted to learn a speckle suppression mechanism from partial volume $v_i$ and a random noise vector $z_i$ to a target speckle-suppressed cross-section $y_i$; $G:\{v_i, z_i\}\rightarrow y_i$.
$G$ is the generator, that is trained to produce speckle-suppressed cross-sections that cannot be differentiated from $y_i$ by an adversarial discriminator $D$.

The objective function of our network is defined as
\begin{align}
    \label{objective function}
    \mathcal{L}_\text{cGAN}(G, D)=\mathbb{E}_{v_i, y_i}\left[\log{D(v_i, y_i)}\right]+ \mathbb{E}_{v_i, z_i}\left[\log{\{1-D(v_i, G(v_i, z_i))\}}\right],
\end{align}
where $\mathbb{E}$ is an ensemble average and $G$ tries to minimize this objective function against an adversarial $D$ that tries to maximize it, i.e., 
\begin{align}
    G^{\ast}= \arg\min_{G}\left\{\max_{D}\left[\mathcal{L}_\text{cGAN}(G, D)+\lambda \mathcal{L}_{L1}(G)\right]\right\},
\end{align}
where
\begin{align}
    \mathcal{L}_{L1}(G)=\mathbb{E}_{v_i, y_i, z_i}[\Vert y_i-G(v_i, z_i)\Vert_{1}].
\end{align}

\begin{figure}
    \centering
    \includegraphics[width=1\columnwidth]{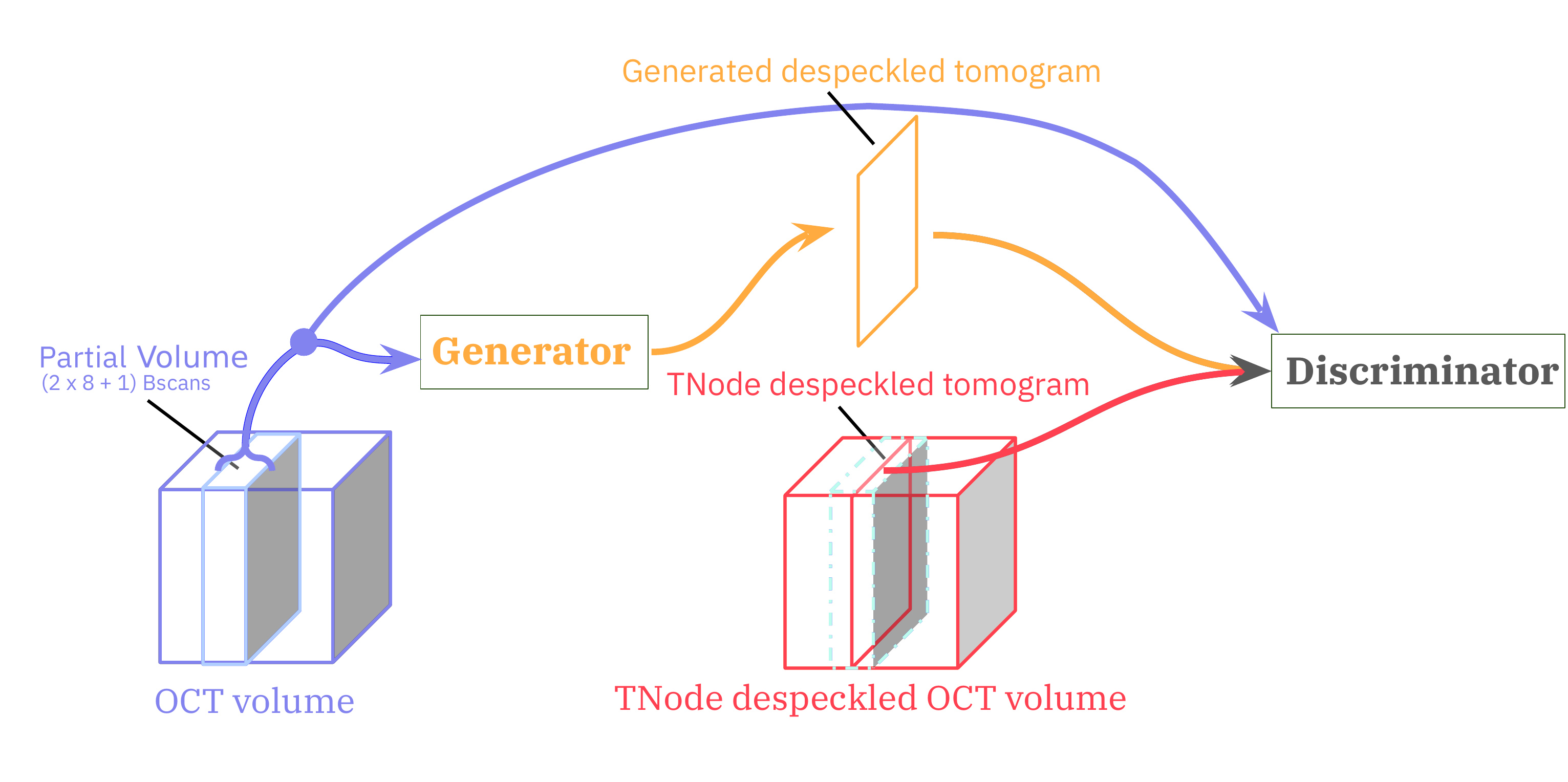}
    \caption{In DL-TNode-3D, a partial OCT volume is given as input to the generator to learn to despeckle tomograms with the help of a discriminator, making use of the given ground-truth speckle suppressed tomogram produced by TNode.}
    \label{fig:processing pipeline}
\end{figure}

Our cGAN network is a modified Pix2Pix network~\cite{isola2017image}, which consists of a generator and a discriminator, as illustrated in Fig.~\ref{fig:processing pipeline}.
The generator we have used in this study is a U-Net with skip connections~\cite{ronneberger2015u}.
The discriminator is a convolutional patchGAN classifier that penalizes at the scale of patches, as illustrated in Fig.~\ref{fig:architecture}.

\subsubsection{Generator}
Our U-Net takes a partial OCT volume ($v_i$) as an input and consists of a series of convolutions and pooling layers, as an encoder, followed by a series of deconvolution and upsampling layers, as a decoder.
In the encoder path, the input arrays are gradually downsampled in $zx$ by half (number of pixels in each dimension) in each layer, while at the same time doubling the number of filter banks.
In contrast, in the decoder path, the input feature maps are upsampled in $zx$ by a factor of two in each layer, while decreasing the number of filters in half.  
The U-Net architecture also consists of `skip' connections, which allows the network to access the information from earlier layers, which otherwise might be lost due to the vanishing gradient problem~\cite{ronneberger2015u}.
The generator loss penalizes the network if the generated speckle-suppressed cross-section is different from the targeted speckle-suppressed cross-section.
The generator loss was defined by a combination of entropy loss and the $L_1$ loss as in the work by Isola et al.~\cite{isola2017image}.
$L_1$ loss allows the generated speckle-suppressed image, $\hat{y_i}$, to be structurally similar to the targeted speckle-suppressed image $y_i$.
As in the work by Isola et al.~\cite{isola2017image}, we provide the noise vector $z_i$ as an input in the form of dropout, applied on several layers of our U-Net generator at both training and testing time.

\subsubsection{Discriminator}
Our discriminator, patchGAN, enables penalization at a patch level to quantify granular details in the generated images.
It consists of a series of blocks tapering down to a desired patch level classification layer, each block consists of a convolution layer, batch normalization, and leaky ReLU as in~\cite{isola2017image}.
We compute discriminator loss in each training step; discriminator loss is a combination of real and fake loss. 
The real loss was a binary cross-entropy loss of patchGAN output for a provided ground-truth speckle-suppressed tomogram $y_i$ and a matrix of ones, while the fake loss was a binary cross-entropy loss of patchGAN output for a generated speckle-suppressed tomogram $\hat{y_i}$ and a matrix of zeros.
Hence, if the generator produced a speckle-suppressed tomogram that matches the provided ground-truth speckle-suppressed tomogram, the real loss will be equal to the fake loss.

\subsection{Training data preparation}
To create a training pair, a random B-scan was first selected from the TNode-processed volume as a ground truth target image, $y_i$, together with 16 adjacent B-scans ($x_{i-8}$ to $x_{i+8}$), and the same B-scan from the raw OCT volume as an input partial volume $v_i$ (17 B-scans in total).
We trained our network using 300 partial volumes that were randomly selected from 3 datasets (100 partial volumes from each dataset).
The logarithmic intensity values in single precision were loaded into Python and normalized to the uint16 range with defined limits described below. Before being fed into the neural network, the data was converted into TensorFloat-32 values (19-bit precision) to leverage the improved performance of modern GPUs compared to traditional 32-bit single-precision data.
We used conventional data augmentation strategies to increase the diversity in our training dataset.
In each training step, we changed the contrast dynamically by varying the lower and upper limits of the uint16 representation.
The lower limit of the contrast range was set to the noise floor of the dataset, and a random value drawn from a uniform distribution ([0, 10]~dB) was added to the lower limit of the contrast range. 
Similarly, the upper limit of the contrast range was set to an average value computed from the volume of interest(11 $\times$ 11 $\times$ 3~px$^3$) centered around the maximum value in the input partial volume, and a random value drawn from a uniform distribution ([-15, 1]~dB) was added to the upper limit of the contrast range.
In the next step, intensity values lesser than or equal to the lower limit of the contrast range were set to 0, and the intensity values equal to or greater than the upper limit of the contrast range were set to 65535; then, all the remaining intensity values were linearly scaled to the interval of [0 65535].
In addition to this data augmentation step, we randomly flipped, rotated, and performed random crop and resize on the training pairs to simulate different geometric orientations.
We normalized the training pairs from -1 to 1 before we gave them as inputs to our network. 

For a given input partial volume, $v_i$, our generator, $G$, generated an estimate of the speckle-suppressed cross-section, $\hat{y_i}$.
The discriminator received two inputs: first, the partial volume $v_i$ and the generated speckle-suppressed cross-section $\hat{y_i}$; second, the partial volume $v_i$ and the ground-truth speckle-suppressed cross-section $y_i$.
In the next step, we computed the generator loss and the discriminator loss and optimized them using an Adam optimizer~\cite{kingma2014adam}.
The network was trained on an NVIDIA RTX 5000 with 24 GB of memory.
We trained both networks for 200 epochs or until the discriminator loss approached $\approx 2\log{2}$.

\subsection{Experiments}
\begin{figure}
    \centering
    \includegraphics[width=\columnwidth]{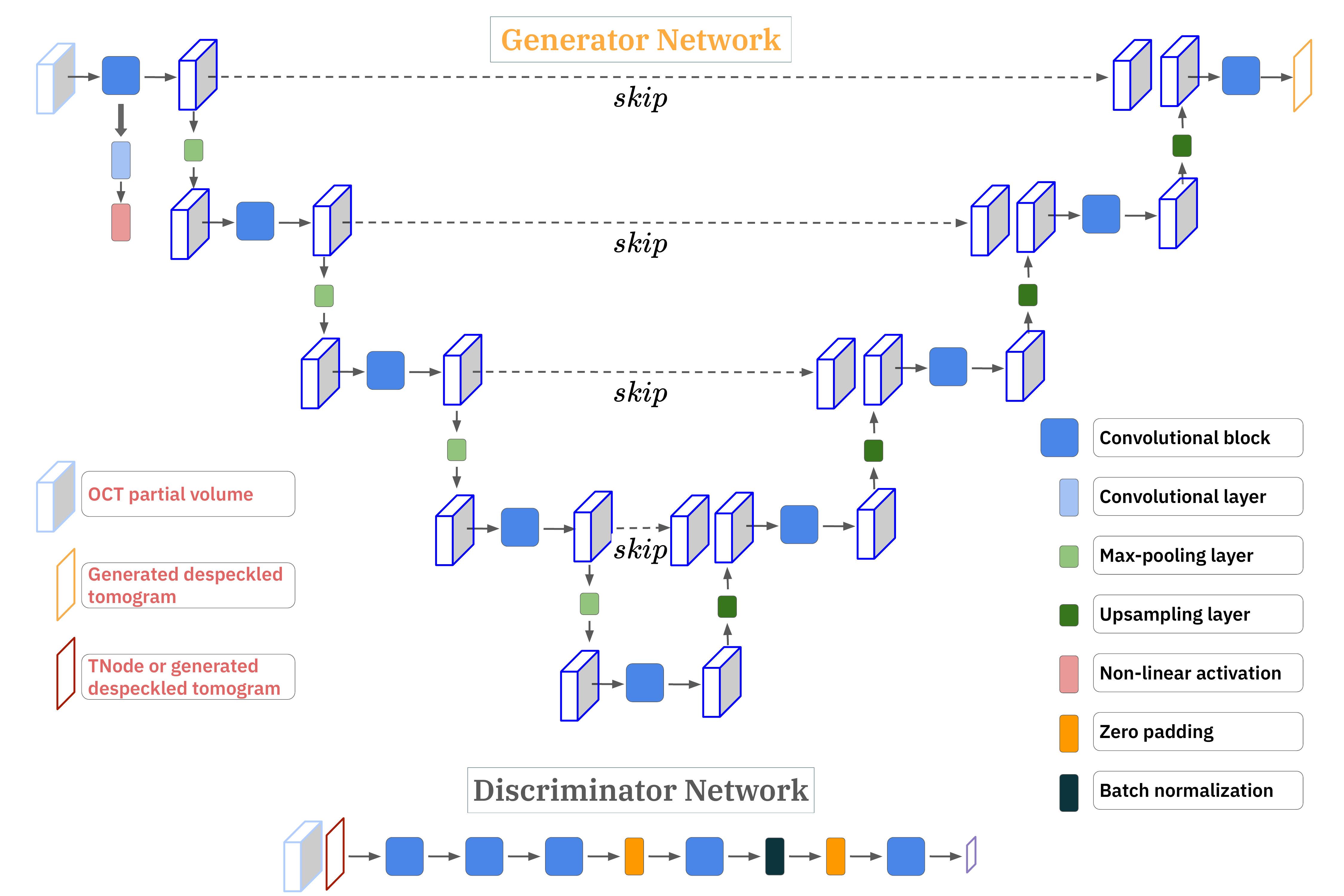}
    \caption{cGAN architecture of our volumetric speckle suppression network; U-Net as a Generator, convolutional patchGAN classifier as a Discriminator.}
    \label{fig:architecture}
\end{figure}

To determine the best architecture for our network, we conducted experiments by changing the number of convolutional blocks in the generator and changing the number of convolutional blocks and the output size of the discriminator with different training and testing datasets.
We converged to the network shown in Fig.~\ref{fig:architecture}, which gave us the best results in terms of time efficiency and accuracy.
The number of A-lines and depth samples of OCT volumes used in this study were less than 1024. Because the search window size used for TNode processing in the out-of-plane direction was 17~px in size, we chose an input partial volume size in our network of 1024~$\times$~1024~$\times$~17~px$^3$.
We used a U-Net with the encoder composed of 5 convolutional blocks with 256, 512, 1024, 2048, and 2048 filter banks, which downsample the input partial volume of size 1024 $\times$ 1024 $\times$ 17~px$^3$ to 32 $\times$ 32 $\times$ 2048~px$^3$.
We used 256, 512, 1024, and 2048 filter banks in respective convolutional blocks in the encoder part. 
We padded with zeros the input partial volumes of size less than the specified input partial volume size in the fast and slow axes. 
Our discriminator consisted of 3 convolutional blocks with 512, 1024, and 1024 filter banks followed by a zero-padding layer and another convolution layer with 2048 filter banks followed by batch normalization, and zero-padding layers, and the last layer was a fully connected layer that outputs a patch-wise classification matrix with size 126 $\times$ 126 corresponding to the classification of patches of size $\approx$ 8 $\times$ 8.

To compare and contrast with the current two-dimensional approach in deep learning despeckling, where a speckle-suppressed B-scan is learned from its corresponding raw B-scan, we also trained our network using solely the central raw B-scan of each subvolume as the input, in combination with the same ground truth target image obtained using the TNode of the full subvolume.
We herein refer to this approach as cGAN-2D, while DL-TNode-3D refers to our volumetric despeckling approach using partial volumes.
We used the same generator and discriminator architecture and data augmentation strategies for cGAN-2D and the same datasets and tissue types for the training.
Detailed comparison of DL-TNode-3D and cGAN-2D is discussed in Sec.~\ref{sec:resultsanddiscussion}.

\subsection{Evaluation metrics}
In this study, the following metrics were used for a quantitative analysis of our network performance:
\begin{itemize}
    \item \textbf{Peak-signal-to-noise ratio (PSNR):} measures the quality of our framework generated speckle-suppressed OCT volume compared to ground-truth speckle-suppressed volume obtained using TNode. 
    \begin{equation*} \text{PSNR} = 10\log_{10}\left(\frac{I_{\max}^2}{\text{MSE}}\right),
    \end{equation*}
    where $\text{MSE} =\sum_{p=1}^{D}( v_p-\hat{v_p})^2$ is the mean square error between the ground truth voxels $v_p$ and the estimated voxels $\hat{v_p}$, $D$ is the total number of voxels in the volume, and $I_{\max}$ is the maximum value possible. MSE computes the cumulative error between the ground truth and the estimated speckle-suppressed OCT volume obtained using our framework. We compute PSNR using 16-bit OCT intensity volumes, hence $I_{max}$ would be $2^{16}-1$. PSNR quantifies the quality of the generated speckle-suppressed OCT volume, a higher PSNR indicates a better quality, as it means that the generated speckle-suppressed OCT volume is closer to the original TNode processed volume in terms of voxel values, therefore less distorted or noisy.  
    
    \item \textbf{Contrast-to-noise ratio (CNR):} measures the contrast between two tissue types.  
    \begin{equation*}\text{CNR} =\frac{|\mu_{\text{t}_2}|-|\mu_{\text{t}_1}|}{\sqrt{\sigma_{\text{t}_2}^2+\sigma_{\text{t}_1}^2}},\end{equation*}
    where $\mu_{\text{t}_1}$ and $\mu_{\text{t}_2}$ are the mean of tissue type \#1 and tissue type \#2 respectively.
    Similarly, $\sigma_{\text{t}_1}$ and $\sigma_{\text{t}_2}$ are the standard deviations of tissue type \#1 and tissue type \#2 respectively. 
    We have computed the CNR between two tissue samples of interest on the speckle-suppressed OCT volume obtained using our framework and compared it with the CNR of the corresponding ground-truth speckle-suppressed OCT volume. The closer the CNR computed on the speckle-suppressed OCT volume obtained using our network to the CNR computed on the ground-truth speckle-suppressed OCT volumes indicates that our framework preserved the contrast between different tissue types as in the ground-truth speckle-suppressed OCT volume.
    
    \item \textbf{Structural similarity index (SSIM) and multi-scale-SSIM (MS-SSIM):} are image quality assessment methods that assess the similarity between corresponding patches of two images ~\cite{wang2004image, wang2003multiscale}.
    The SSIM is computed using three components, namely the luminance ($I$), the contrast ($C$) and the structural ($S$), and they are defined as
    \begin{align*} I(y,\hat {y}) = \frac {2~\mu _{y} \mu _{ \hat {y} } + C_{1} }{\mu _{y}^{2} + \mu _{\hat {y}}^{2}+C_{1} }, \;
    C(y,\hat {y}) = \frac {2\sigma _{y} \sigma _{ \hat {y} } + C_{2} }{ \sigma _{y}^{2} + \sigma _{\hat {y}}^{2}+C_{2} }\ 
    \mathrm{and}\ 
    {{S(y,\hat {y})= \frac {\sigma _{y\hat {y}} + C_{3} }{ \sigma _{y} \sigma _{\hat {y}} + C_{3} },} } \end{align*}
    where $\mu_{\hat{y_i}}$ and $\sigma_{\hat{y_i}}$ are the mean and the standard deviation of the generated speckle-suppressed cross-section ${\hat{y_i}}$, respectively; $\mu_{y_i}$ and $\sigma_{y_i}$ are the mean and the standard deviation of the label image $y_i$, respectively; $\sigma_{y_{i}\hat{y_i}}$ denotes the cross-covariance between ${\hat{y_i}}$ and $y_{i}$; $C_{1}$, $C_{2}$ and $C_{3}$ are small positive values used to avoid numerical instability.
    $\text{SSIM}(y_i,\hat{y_i})$ is the product of these three components,
    \begin{equation*} \text{SSIM}(y,\hat {y}) = I(y,\hat {y})^\alpha C(y,\hat {y})^\beta S(y,\hat {y})^\gamma,  \end{equation*}
    where $\alpha$, $\gamma$ and $\beta$ are exponent weights for the luminance, contrast and structural components.
    SSIM is measured using a fixed patch size, which does not capture complex variations between the input two images to assess the similarity. MS-SSIM considers the input patches that are iteratively downsampled by a factor of two with low-pass filtering, with scale $j$ denoting the original images downsampled by a factor of $2j-1$ and given as 
    \begin{equation*}\text{MS-SSIM} (y, \hat {y}) = I_{M}(y,\hat {y})^{\alpha M}\prod _{j=1}^{M}C_{j}(y,\hat {y})^{\beta _{j}}S_{j}(y,\hat {y})^{\gamma _{j}}. \end{equation*}
    Note that these equations are valid regardless of the dimensionality of the images being compared: for use in volumes, we compute $I(y,\hat {y})$, $C(y,\hat {y})$, and $S(y,\hat {y})$ over 3D-patches instead of 2D-patches.

\end{itemize}
We used MATLAB functions \textit{ssim}\footnote{https://www.mathworks.com/help/images/ref/ssim.html} and \textit{multissim}\footnote{https://www.mathworks.com/help/images/ref/multissim.html} with default values for patch size, exponent weights $\alpha$, $\gamma$ and $\beta$ and constants $C_{1}$, $C_{2}$ and $C_{3}$ to compute both volume SSIM and MS-SSIM respectively in this study.
MATLAB \textit{ssim} function also returns the local SSIM value for each voxel.
Using these local SSIM maps, we evaluated the statistical significance between DL-TNode-3D and cGAN-2D using the Student's t-test. 

\section{Results and Discussion}
\label{sec:resultsanddiscussion}

\begin{figure}
    \centering
    \includegraphics[width=\columnwidth]{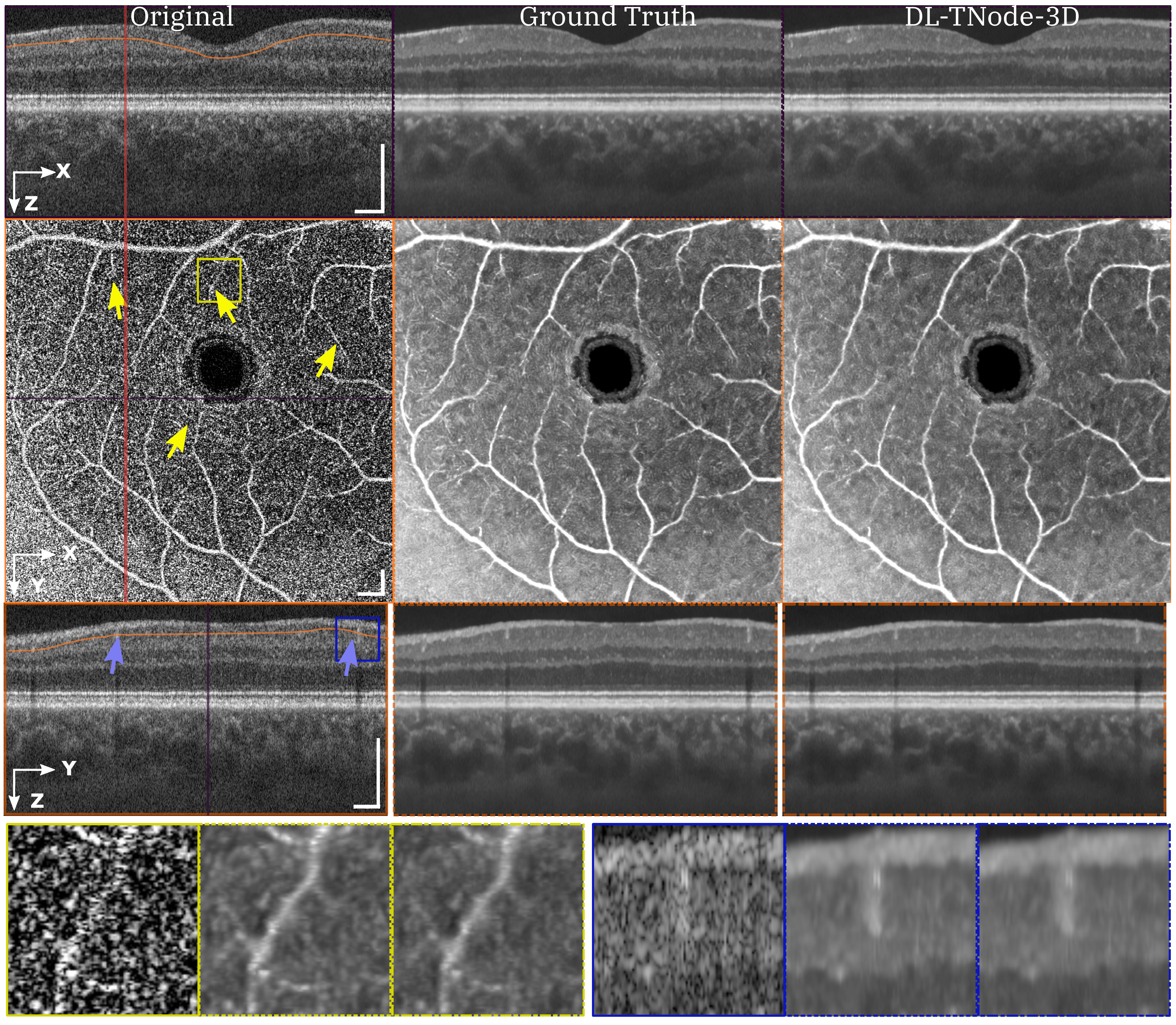}
    \caption{Orthogonal views of tomograms before (a) and after despeckling using TNode (i.e. Ground Truth) (b) and DL-TNode-3D (c). DL-TNode-3D produces OCT volumes close to the ground truth without any visible artifacts along the out-of-plane axis, $y$. $z$ is the depth and $x (y)$ is the fast- (slow-) scan axis direction. Yellow arrows indicate small capillaries that are preserved after despeckling with both TNode and DL-TNode-3D. Scale bars = 0.5~mm.}
    \label{fig:Retina}
\end{figure}

Figure~\ref{fig:Retina} shows representative results for a retinal volume using the ophthalmic OCT system, when trained with three different OCT volumes consisting of distinct fields of view of the retina of two individual subjects.
Our network produced OCT volumes that closely resembled TNode volumes: however, the cGAN processed the entire volume of size (448 $\times$ 818 $\times$ 808 px$^3$) in just 2 minutes (40~ms per B-scan), as opposed to the 4 hours required for TNode on an NVIDIA RTX 5000 with 24 GB of GPU memory, which is an \emph{improvement of two orders of magnitude in processing time}. 
We quantified the similarity between the ground truth and our network-produced tomograms using the volume SSIM and MS-SSIM; for this example, the respective values were 0.988 and 0.996.
To the best of our knowledge, our network produced speckle-suppressed tomograms more similar to the ground truth---as measured by SSIM and MS-SSIM metrics, and regardless of the method used for ground-truth generation---than any other method in the literature.
Our network enhanced the contrast between the layers, similar to ground-truth TNode speckle-suppressed cross-sections while preserving small structures. For instance, the small capillaries, marked with yellow arrows in Fig.~\ref{fig:Retina}, became much clearer and easier to identify.

\begin{figure}
    \centering
    \includegraphics[width=\columnwidth]{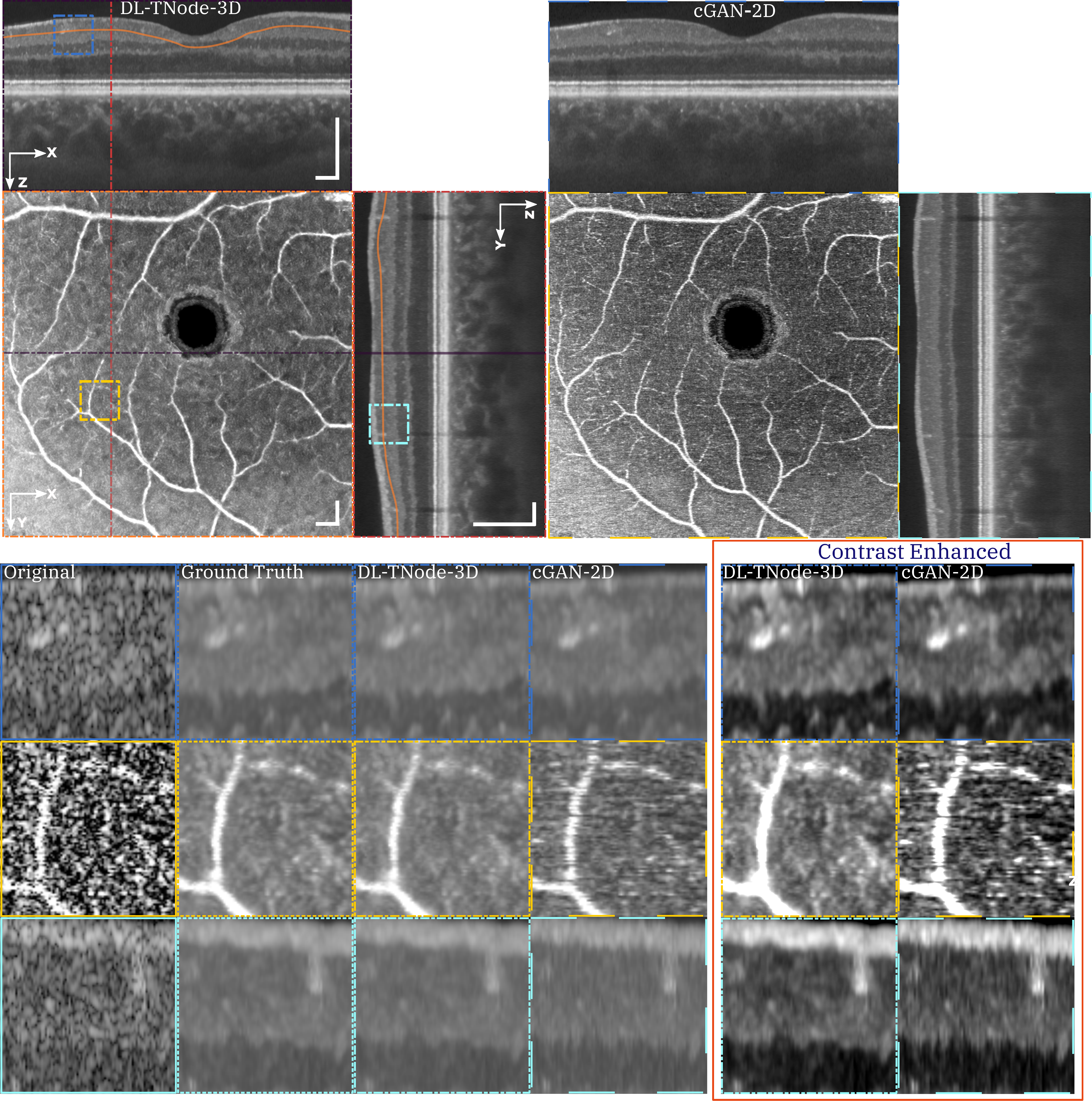}
    \caption{Comparison of speckle reduction using DL-TNode-3D and cGAN-2D for the retinal volume in Fig.~\ref{fig:Retina}. Orthogonal views where $z$ is depth and $x (y)$ is the fast- (slow-) scan axis direction. DL-TNode-3D produces OCT volumes close to the ground truth without any visible artifacts along the out-of-plane axis, $y$. Contrast-enhanced boxes show superior speckle suppression ability of DL-TNode-3D compared to cGAN-2D, which exhibits high-frequency artifacts along the slow-scan axis. Scale bars = 0.5~mm.}
    \label{fig:Retina_comparison}
\end{figure}

Figure~\ref{fig:Retina_comparison} shows a comparison of speckle reduction with cGAN-2D, the case in which we trained our network using the single raw B-scan of interest as the input---instead of a partial volume---and its corresponding TNode processed B-scan as the target image.
Limiting the learning and inference process to 2D processing with cGAN-2D produced a decrease in the quality of the results (similarity metrics were 0.947 and 0.976 for volume SSIM and MS-SSIM, respectively).
It is clear from the results in Fig.~\ref{fig:Retina_comparison} that 2D processing produces high-frequency artifacts along the slow-scan axis direction, and the quality of despeckling is qualitatively and quantitatively inferior to 3D processing.
This demonstrates that the excellent performance of DL-TNode-3D is due to the use of partial volumes for network training and inference, which contain more structural information than individual B-scans. 
We observed similar DL-TNode-3D performance in tomograms acquired using the two other OCT systems, for which the network had been trained independently.

\begin{figure}
    \centering
    \includegraphics[width=\columnwidth]{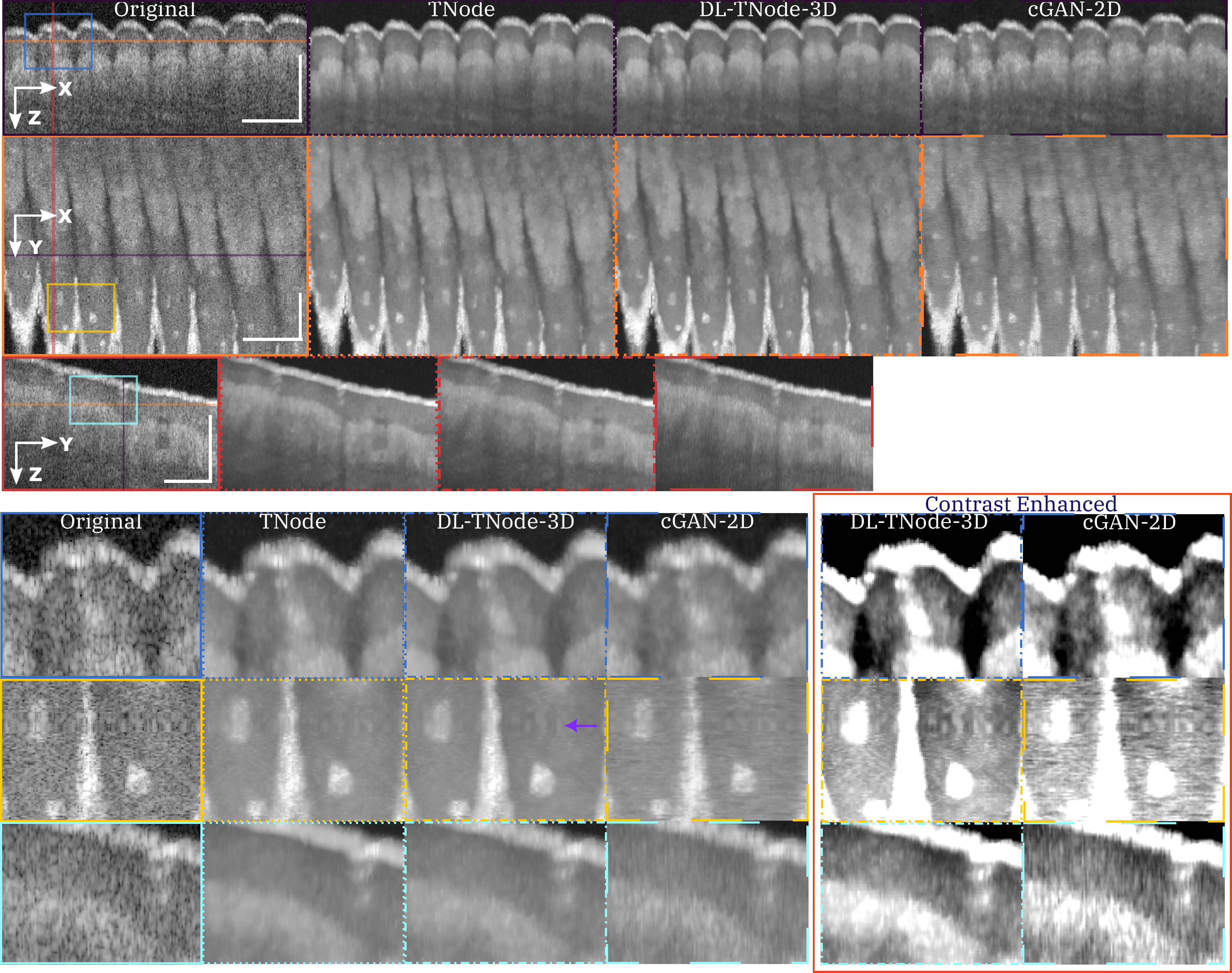}
    \caption{Comparison of speckle reduction using DL-TNode-3D and cGAN-2D for the ventral finger skin volume acquired using VCSEL system. Orthogonal views where $z$ is depth and $x (y)$ is the fast (slow) axis direction. DL-TNode-3D produces OCT volumes that match the ground truth without any visible artifacts along the out-of-plane axis, $y$. Purple arrow in the en face view indicates a motion artifact. Contrast-enhanced boxes show superior speckle suppression ability of DL-TNode-3D compared to cGAN-2D, which exhibits high-frequency artifacts along the slow-scan axis. Scale bars = 0.5~mm.}
    \label{fig:VCSEL_example}
\end{figure}

Figure~\ref{fig:VCSEL_example} shows representative results for a ventral finger skin volume (tissue type not part of training data) using the VCSEL-based OCT system when trained with three OCT volumes consisting of the nail bed, chicken heart, and chicken leg tissues.
The similarity between the ground truth and our network-produced tomograms using the volume SSIM and MS-SSIM for this example were 0.968 and 0.993, respectively.
In contrast, testing with cGAN-2D resulted in a decrease in the quality of the results with volume SSIM and MS-SSIM 0.960 and 0.986, respectively.
Furthermore, cGAN-2D processing produced high-frequency artifacts along the slow-axis scan direction, and the quality of speckle suppression is qualitatively inferior to DL-TNode-3D.
This network gave similar results when we tested on dorsal finger skin volume with volume SSIM and MS-SSIM 0.988 and 0.996, respectively.  
This experiment shows that our network is generalizable for tissue types that are not part of the training data.
Speckle reduction using our network enhanced the contrast between different ventral and dorsal finger skin layers.
Sweat ducts can be identified easily on our network speckle-suppressed cross-sections. 

\begin{figure}
    \centering
    \includegraphics[width=\columnwidth]{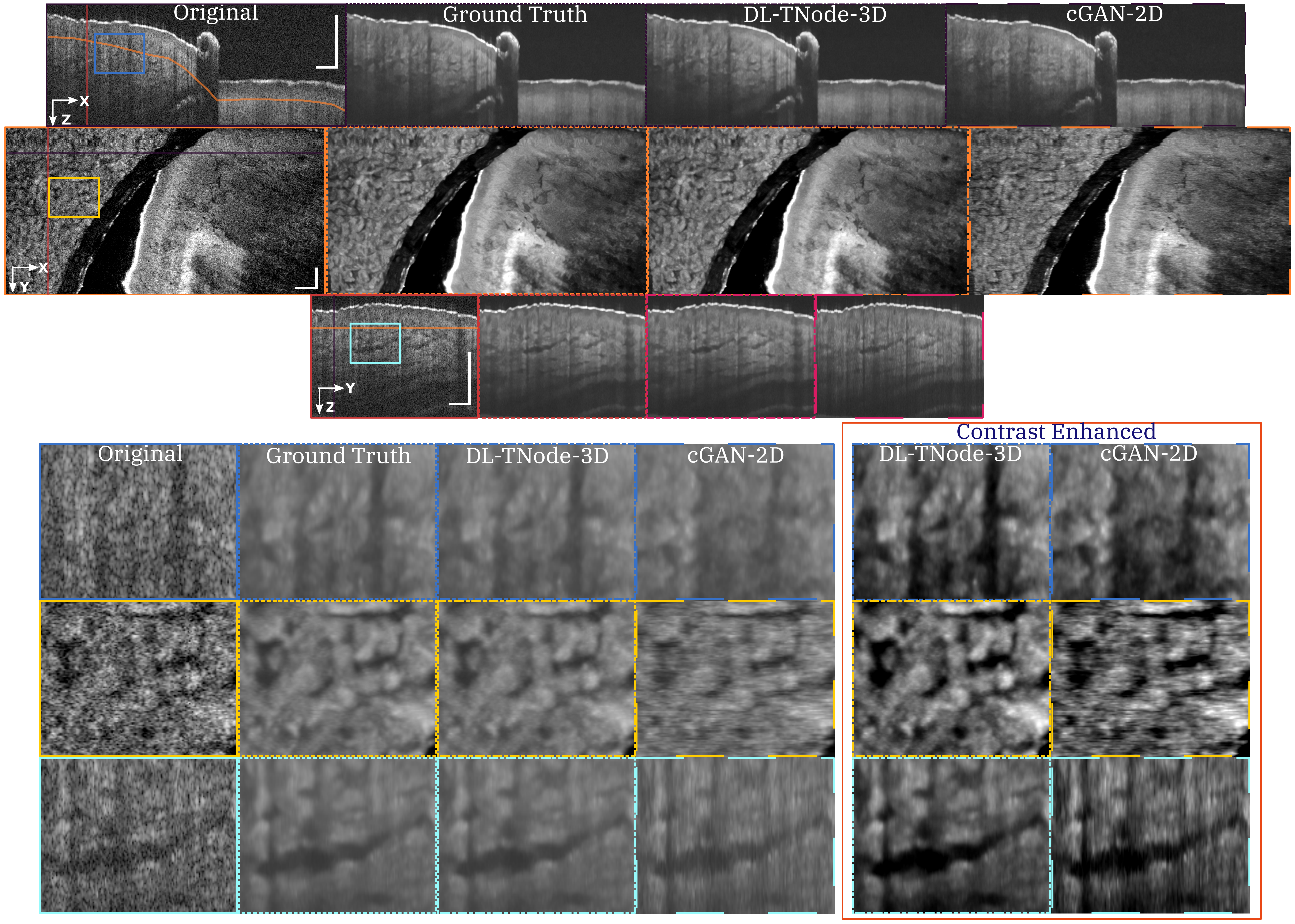}
    \caption{Comparison of speckle reduction using DL-TNode-3D and cGAN-2D for the ventral nail bed volume acquired using the polygon-based system. Orthogonal views where $z$ is depth and $x (y)$ is the fast (slow) axis direction. DL-TNode-3D produces OCT volumes close to the ground truth without any visible artifacts along the out-of-plane axis, $y$. Contrast-enhanced boxes show superior speckle suppression ability of DL-TNode-3D compared to cGAN-2D, which exhibits high-frequency artifacts along the slow-scan axis. Scale bars = 0.5~mm.}
    \label{fig:JianSystem_example}
\end{figure}

Similarly, Fig.~\ref{fig:JianSystem_example} shows representative results for a nail bed volume using the polygon-based OCT system when trained with three OCT volumes consisting of chicken leg (2$\times$) and dorsal finger skin (1$\times$), respectively.
The similarity between the ground truth and our network-produced tomograms using the volume SSIM and MS-SSIM for this example were 0.989 and 0.996, respectively.
In contrast, testing with cGAN-2D resulted in a decrease in the quality of the results with volume SSIM and MS-SSIM 0.943 and 0.983, respectively.
It is clear from the results in Fig.~\ref{fig:JianSystem_example} that 2D processing produces high-frequency artifacts along the slow-axis scan direction, and the quality of despeckling is qualitatively inferior to 3D processing.

The quality metrics for the ophthalmic, VCSEL, and polygon systems are summarized in Table~\ref{tab:qualitymetrics}.
\begin{table}
 \caption{Quantitative evaluation of DL-TNode-3D and cGAN-2D for 3 OCT systems.
 DL-TNode-3D is our method, which uses the volumetric information for speckle suppression whereas cGAN-2D uses only 2D (depth-fast axes) information for speckle suppression.
 CNR values that are nearest to ground truth volume CNR are highlighted in bold.
 The highest values of PSNR, SSIM, and MS-SSIM are highlighted in bold.
 CNR: contrast-to-noise ratio, PSNR: peak-signal-to-noise ratio; SSIM: structural similarity index; MS-SSIM: multi-scale structural similarity index.}
    \begin{center}
        \begin{NiceTabular}{c|c|cccc} \hline
        \textbf{OCT System}  &  \textbf{Trained Model} & CNR & PSNR (dB) $\uparrow$ & SSIM $\uparrow$ & MS-SSIM $\uparrow$\\ \hline \hline
         Ophthalmic & Ground truth & 1.193 & - & - & - \\
        & cGAN-2D & \textbf{1.188} & 34.726 & 0.943 & 0.976 \\ 
        & DL-TNode-3D & 1.302 & \textbf{38.076} & \textbf{0.988}  & \textbf{0.996}  \\ \hline
        VCSEL & Ground truth & 1.623  & - & - & - \\
         & cGAN-2D & \textbf{1.612} & 37.175 & 0.954  & 0.982 \\ 
        & DL-TNode-3D & 1.675 & \textbf{41.095} & \textbf{0.978} & \textbf{0.994}\\ \hline
        Polygon & Ground truth & 1.394 & - & - & - \\
        & cGAN-2D & 1.525 & 36.876 & 0.949 & 0.985 \\ 
         & DL-TNode-3D & \textbf{1.505} & \textbf{40.654} & \textbf{0.988} & \textbf{0.996} \\ \hline
        \end{NiceTabular}
    \label{tab:qualitymetrics}
    \end{center}
\end{table}
It is evident that considering the out-of-plane information for speckle suppression using our DL-TNode-3D framework produced volumes most similar in terms of volume PSNR, SSIM and MS-SSIM to the ground truth volumes compared to using only B-scans in cGAN-2D.
The p-value was $ < 10^{-6}$ between the SSIM of DL-TNode-3D and cGAN-2D for all three OCT systems, which shows us that DL-TNode-3D performance is superior to cGAN-2D with unconditional distinction.
cGAN-2D produced volumes with CNR close to the ground truth CNR for models trained on data from the ophthalmic and VCSEL systems but not from the polygon-based system. However, they are corrupted with high-frequency artifacts in the out-of-plane direction in all cases. 
Figure~\ref{fig:Retina_comparison} illustrates the potential use of our network for glaucoma monitoring as it increases the contrast between retinal layers (nerve fiber layer, ganglion cell layer and inner plexiform layer) for which longitudinal monitoring of thickness is important~\cite{banister2016can,mwanza2015retinal}.
We expect that automated segmentation routines will perform more accurately on despeckled volumes, improving the tracking of layer thicknesses. Even for deep-learning segmentation routines, we expect that the use of despeckled data to generate more accurate training data will be beneficial.

Our results show that our network trained with only three OCT volumes of readily available tissue can produce despeckled volumes that replicate the efficient suppression of speckle and
preservation of tissue structures with dimensions approaching the system resolution known from TNode, while being two orders of magnitude faster.
Because DL-TNode-3D relies on an all-software approach for training, it can be easily re-trained and deployed in virtually any OCT system. The updated TNode code for generating the training data and the source code for our neural network are available in \href{https://github.com/bhaskarachintada/DLTNode.git}{GitHub repository},~Ref.~\cite{Chintada2023DLTNode}).
\section{Conclusion}
In this work, we presented a despeckling cGAN framework to utilize volumetric information for speckle suppression in OCT.
Our framework was trained using partial OCT volumes as input and TNode speckle-suppressed tomogram as targeted output using U-Net as a generator and patchGAN as the discriminator.
We trained and tested our framework using 300 partial volumes randomly drawn from three OCT volumes (100 partial volumes from each OCT volume) from three OCT systems separately.
Our network produced volumes that approximate the ground-truth despeckled volumes with unprecedented fidelity, for models trained on all three OCT systems.
Additionally, our framework is two orders of magnitude faster than TNode and reaches near real-time performance at $\sim$ 20 fps.
\pagebreak
\section*{Supporting material} See Supplement for supporting content.
\begin{backmatter}
\section*{Funding}
National Institutes of Health (P41 EB015903, R01 EB033306, K25 EB024595), Department of Defense through the Military Medical Photonics Program (FA9550-20-1-0063), and Universidad EAFIT (11100252021).
\section*{Acknowledgments}
We gratefully acknowledge Dr. Neerav Karani, Computer Science and Artificial Intelligence Laboratory, Massachusetts Institute of Technology, Cambridge, MA, USA for discussions on cGAN.

\section*{Disclosures}
The authors declare that there are no conflicts of interest related to this article.
\end{backmatter}

\bibliography{bibliography}
\end{document}